# Large Instrument Development for Radio Astronomy

## Astro2010 Technology Development White Paper

From
Observatory Technical Council, National Radio Astronomy Observatory
Charlottesville, VA; Green Bank, WV; Socorro, NM, Tucson, AZ

Principal Author: J. Richard Fisher (OTC Chair), 434-296-0206, rfisher@nrao.edu

Co-Authors: Richard F. Bradley, Walter F. Brisken, William D. Cotton, Darrel T. Emerson, Anthony R. Kerr, Richard J. Lacasse, Matthew A. Morgan, Peter J. Napier, Roger D. Norrod, John M. Payne, Marian W. Pospieszalski, Arthur Symmes, A. Richard Thompson, John C. Webber

Abstract

This white paper offers cautionary observations about the planning and development of new, large radio astronomy instruments. Complexity is a strong cost driver so every effort should be made to assign differing science requirements to different instruments and probably different sites. The appeal of shared resources is generally not realized in practice and can often be counterproductive. Instrument optimization is much more difficult with longer lists of requirements, and the development process is longer and less efficient. More complex instruments are necessarily further behind the technology state of the art because of longer development times. Including technology R&D in the construction phase of projects is a growing trend that leads to higher risks, cost overruns, schedule delays, and project de-scoping. There are no technology breakthroughs just over the horizon that will suddenly bring down the cost of collecting area. Advances come largely through careful attention to detail in the adoption of new technology provided by industry and the commercial market. Radio astronomy instrumentation has a very bright future, but a vigorous long-term R&D program not tied directly to specific projects needs to be restored, fostered, and preserved.

# Large Instrument Development for Radio Astronomy

## Introduction

The size, cost, and sophistication of radio astronomy instrumentation have grown enormously in its relatively short history and must continue to grow. New discoveries and the solutions to fundamental problems will require continued technological advances. However, greater technical sophistication does not necessarily mean that greater complexity is required. We argue in this white paper that strategies must be adopted for reducing complexity and that the trend toward very large international instruments could be counterproductive. The advantages of scale and shared infrastructure will be more than offset by inefficiencies of management complexity and scientific and technical compromises. Most of the fruits of collaboration can and do occur at the individual and institutional level in ways that already work very well. The advantages of competing approaches and higher risks permitted at smaller scale must be at the heart of radio astronomy for the indefinite future, not just as precursors to projects that consume most of our resources.

One prevailing assumption is that, through a combination of pooling of resources and being much cleverer, we can create one or two orders of magnitude increases in sensitivity and fields of view for the cost of a moderate size space mission. In fact, there is only a modest amount of untapped cleverness or new technology waiting to be employed that will make the more visionary advances possible in the coming decade. As prosaic as they appear, the current front-line instruments in radio astronomy are reasonably well optimized for cost and scientific capability. To a first approximation, ten times as much money will buy ten times as much collecting area, all else being equal. The same factor of ten in cost will not buy ten times the collecting area, ten times the field of view, and maintain the wide frequency coverage, instantaneous bandwidth, system temperatures, and spatial resolution of the best current instruments. The total performance to cost ratio is not poised to make a great leap forward as is implied in our grander proposals, Moore's law notwithstanding. Trade-offs can be optimized for different scientific problems, *but not all in the same facility*.

## Transformational technology is rare and starts small.

The vast majority of new technology employed by radio astronomy is developed in industries with far greater resources for research and development. There are notable exceptions to this general rule, such as SIS mixers, but these exceptions are rare. Even the transformative technology of cryogenic amplifiers using High-electron-mobility Field-Effect Transistors (HFETs), resulted from a brilliant recognition that a particular run of HFET's developed for other purposes work extremely well when cooled. The best HFET chips employed in nearly all receivers over the past two decades come from a very small number of fortuitously good semiconductor wafers.

Technological advances in radio astronomy are largely the result of careful attention to detail and the adoption of new technology from the consumer and industrial market. What we are good at is adapting technology for our purpose, but, on average, the technology in use by today's best radio telescopes is 10 to 20 years old. Larger pieces of instrumentation tend to contain older technology. For some things, such as antennas, where the technology changes slowly, there is only minor disadvantage to age, but for subsystems such as signal processing the age penalty is enormous. Size and complexity are the enemies of signal processing cost over the lifetime of an instrument. This applies to all aspects of the developments cycle - specification, hardware design, software design, debugging and verification, and user interfaces. Efforts to segregate scientific requirements into categories with similar instrumentation needs can pay dividends compared to current practices and should be a major consideration in new instrument developments.

“Shared resources” is a fuzzy concept whose advantages we take on faith. This tends to lock us into the circular assumptions that: (1) we can afford only one big instrument so (2) it must satisfy as many scientific requirements as possible, which (3) makes it complex, costly, and slow to build, and, therefore (4) we can afford only one every 20 years. A few sub-disciplines, such as pulsar astronomy, have elected to resist this cycle by continuously developing small instruments with lifetimes less than about ten years. The trend in this field is to reduce the development time even further. Pulsar astronomers are more attuned to signal processing technology so they reap the advantages of faster development cycles made possible by improvements in the cost of computation. This strategy could be applied in other areas, but the advantages are not as evident to other scientists. The operations budget of a large instrument should assume the need for aggressive and continuous enhancements with a sustained technology R&D program to reduce the need for infrequent, very expensive, and disruptive major upgrades.

Clearly, something such as a signal correlator for a synthesis array cannot be divided into sub arrays with the same total capability, but serious consideration should be given to the question of whether the same correlator should serve more than two or three octaves of observing frequency. Higher frequencies need more bandwidth, and lower frequencies need more dynamic range. A narrower-bandwidth correlator built for low-frequencies may be built with field-programmable gate arrays (FPGAs) whose flexibility can be turned toward RFI excision and pulsar applications. These functions would only add unnecessary complexity to a wider-bandwidth correlator that must be built for high frequencies with application-specific integrated circuits (ASICs) which require large non-recurring engineering (NRE) costs.

New technologies are often cited as reasons to expect that great advances in capability are on the horizon. Examples include object-oriented programming, MMICs, FPGAs, hydroformed reflectors, and ultra-wideband antennas. Indeed, each of these is a significant advance to be exploited, but each involves system compromises and concurrent advances in related technologies that moderate their advantages to new system capability. In practice, advances such as these require a great deal of spade work to learn and incorporate the new technology. The advantages of scale are also sometimes cited as reasons for optimism that a big new instrument will be less costly per unit performance. The reality is that in radio astronomy parts quantities are orders of magnitude smaller than those in the consumer market, even for the most ambitious proposed systems, so the cost savings of quantity are mostly in amortized NRE, not in mass-manufacturing advantages.

## Different Frequency Ranges Require Different Antennas and Different Sites

Antennas work very well over about one decade of frequency range. Surface and pointing errors erode efficiency of reflector antennas at frequencies above their optimum range. At lower frequencies diffraction effects reduce efficiency and increase spillover noise, particularly in a Cassegrain or Gregorian system where diffraction is most affected by the relatively small subreflector. Above the highest optimum frequency the beam size is smaller, and the antenna weight and wind resistance are greater than necessary for the effective collecting area achieved. Hence, the system is penalized in field of view, and the demands on the pointing system are much higher. Radio astronomers have done remarkably well at squeezing the most out of antennas beyond their optimum frequency ranges in the absence of anything better, but we have forgotten how severe the compromises have been. As we consider radio telescope systems with very large collecting areas, extreme dynamic range demands, and barely affordable costs, efficiency and accuracy compromises become much more costly than with current instruments.

At frequencies above about 10 GHz atmospheric effects are the greatest factors in selecting a site. Below about 1.5 GHz, radio frequency interference (RFI) is the strongest consideration. Above this frequency most strong RFI at any reasonable radio astronomy site comes from satellites from which even very sparsely populated areas are not immune. Locations such as southern Africa and Western Australia are indisputable choices for new low-frequency arrays. At higher frequencies the site choices are more diverse, and factors such as accessibility, longest available baselines, and existing infrastructure will play a larger role in the site selection process.

## How important is a common site?

Proponents of the next generation of radio telescopes cite shared infrastructure and operating resources as a strong reason for selecting a common location for most new construction. If we accept that systems optimization and reduced complexity are good arguments for building different antennas and correlators for different frequency ranges, then the shared infrastructure is fairly basic - roads, power lines, cable trenches, housing, and the like. The total power requirements, communications bandwidth, high-level operations and maintenance personnel will be roughly the same for three substantially different instruments at one site compared to three separate sites. *Attempts to share key components of different instruments, such as correlators, antennas, or data transmission bandwidth will reduce the observing time available for each frequency range*.

Any one instrument will be staffed well above critical mass for a viable intellectual and management environment. Collaborations at the component and subsystem development level do and will happen between people scattered all over the world whenever there is mutual benefit. A light-weight international management system to clear administrative impediments and foster communications will be very helpful, but most of the professional exchange channels have been in place for decades and are very effective. Collaborations will form and dissolve as the needs and opportunities arise.

Another advantage of assigning scientific and technical requirements to different instruments is that fewer organizations need be involved in the details of funding and siting of each instrument. Satisfying the different, sometimes conflicting, regulatory and political requirements of a number of different funding agencies forces expenditure decisions to be made which may not be optimum based on scientific or technical considerations. In addition to leaner management structures, there are advantages to unlocking funding time lines, site selections, and development schedules.

## How important is system temperature?

The best radio astronomy receivers regularly operate with system temperatures that are pretty close to the limits set by atmospheric and cosmic background noise plus the inevitable antenna spillover noise. However, these cryogenic receivers are heavy, big, and relatively expensive, and their bandwidths are generally limited to about 50% of center frequency. The demands of much greater collecting area and fields of view make current receiver designs impractical for many concepts for much larger collecting area and fields of view. Hence, the prevailing design trend is to accept higher system temperature from uncooled receivers or receivers cooled to higher temperatures and to recoup the lost sensitivity with more collecting area or more beams on the sky. This trade-off has not yet been shown to be beneficial in a full design optimization.

The problem is that for a given point source sensitivity the collecting area required is inversely proportional to the system temperature, and for a single aperture the field of view is inversely proportional to collecting area. To recover the field of view, more beams must be created with extra hardware and signal processing power. A synthesis array could add more antennas of the same size to maintain the field of view, but this drives up a large portion of the signal processing costs in proportion to the square of the system temperature. The post-correlation data processing load will also be driven up by the increase in the number of antennas or number of simultaneous beams. The costs of much larger apertures are going to be strongly dependent on the cost of signal processing, so this is not a place where sensitivity can be economically recovered.

In presentations of alternative solutions for building greater collecting area and field of view there is a disturbing tendency to take an incomplete view of the optimization trade-offs, always to the benefit of the particular point of view. Honest assessments of subsystem costs that meet the scientific requirements are usually daunting. This leads to a round of requirements trimming and parameter adjustments to get the total cost below a value believed to be feasible. Every step in this trimming process is biased in an optimistic direction, usually unconsciously. The result is a powerful recipe for cost inflation and design and construction delays when hard reality must be reckoned. Funding cycle times have not increased with increased specification and costing times of larger projects. The result is that few, if any, new large projects have had the time to establish reasonably accurate costs before full-scale funding campaigns begin.

## What is needed and what is possible.

As the size and complexity of instruments have grown there are fewer and fewer individuals who can effectively weigh the scientific and technical trade-offs of a new system design.

Hence, we have evolved into a rather linear process beginning with science requirements followed by system design and implementation. Scientific requirements are generally set by a committee of vested interests. The result is an "anding" of most, if not all individual requirements. Feedback on how various and sometimes conflicting science requirements affect cost and design and construction time is usually based on insufficient information.

Overly ambitious specifications become "goals", which are generally treated as firm specifications in the design process because there is no weighting metric for the transition from requirement to goal. A commercial supplier will ignore a goal for lack of strong financial incentive. Staff engineers treat goals in a variety of ways, depending on their individually perceived weights of the required compromises. They may have no immediate incentive to reduce cost; otherwise the goal would have been dropped in the first place. A variation on the fuzzy "goal" concept is the request to "not design it out." This can be even more costly than goals and an even greater cost driver than the well-defined specifications. If scientists knew that the time between new instrumentation upgrades were shorter than is now typical, the propensity toward squeezing all requirements into one design and modifying requirements in mid-project may be reduced.

Shortcomings in the linear requirements-through-design process are dealt with in several levels of project review with titles such as "conceptual", "preliminary", "critical", "pre-production", and "final". These usually involve a lot of report and presentation preparation and travel. There is definite value in these reviews, but any scientist with a strong interest in instrumentation design will recognize how inefficient the process is and may dread being assigned to one of these review committees. These inefficiencies, as a fraction of total project cost, grow with project complexity for a given level of talent and expertise. An additional danger of excessive reliance on the review "process" is that it can lull a project into a false sense of security. We cannot review the answer to a question that we are not clever enough to ask. Modern radio telescopes are sufficiently close to the state of the art that we should at all times admit the possibility of a surprise. There is no substitute for experience and adequate prototyping.

We believe there are measures we can take to improve this process. They begin with a greater level of trust and communication between scientists and engineers. If an engineer says that there is a certain level of uncertainty and risk in meeting certain requirements, this needs to be taken at face value and built into the management process. Scientists need to be free to communicate the relative weights they place on different requirements with the confidence that stretched goals will be neither ignored nor taken as requirements at any cost.

Everyone needs to understand that the only designs for which accurate costs and production times are available are ones that have been fully prototyped and tested. The most frequent cause of project de-scoping and/or cost overruns is overly optimistic cost and development time estimates based on insufficient prototyping. We should always be pushing the envelope, but deferring R&D time and costs until after a project is funded will cause the project cost to become less and less certain, almost always in the direction of underestimated cost and overly optimistic schedules. Paper studies are not a substitute for full prototyping.

Prototyping is an enormously valuable and under-appreciated skill. Too little prototyping does not sufficiently retire risks and too much wastes time and resources. Initial prototyping is exploratory, and later stages are aimed at reliability and cost control. "Demonstrator" projects are often good examples of poor prototyping. They are often built to convince committees and funding agencies that a project is viable. They spend far too much time on building subsystems that are already well understood, and too few resources are allocated to retiring real risks.

Experts need far more detailed proof that key unknowns are under control than is typically provided by a "demonstrator." Non-experts, even renowned scientists, are easily fooled by an impressive but superficial presentation.

All risks cannot be retired before the beginning of a project, and improvements of the design at any stage of construction are possible, but the zero-contingency budget must be consistent with a high certainty of on-time completion that meets the core specifications. Higher risks must be consistent with a higher contingency. This seems so obvious, but through a combination of naiveté, wishful thinking, and funding desperation, it is being treated far too lightly in the current generation of proposals. Visionaries come from the ranks of those without a sobering failure under their belt.

## How much preliminary R&D is enough and how is it evaluated?

Research and development is reasonably cheap in the grand scheme of research facilities. A minimum, viable, long-term R&D program will very roughly require a budget of two percent of the annual operating costs of existing observing facilities. This would pay for research engineer salaries, technician support, new test equipment, and supplies. One reason that this activity is not being pursued at an adequate level is that this money is discretionary in the short term and hard to protect under tight budget conditions. Another reason is that review panels at funding agencies are requiring closer and closer ties of instrument development to immediate scientific return. Hence, a growing fraction of instrument development funds is going into construction at the expense of more innovative device or system development.

Since good R&D is inherently high risk, deliverables are hard to promise. A perfectly valid result in six months time is that a certain approach will not work, and now we know why. Each step in the process can be a branch point – “this worked, keep going,” or “this didn't work, try something else.” New ideas are being generated all the time, and even perfectly valid approaches may be set aside in pursuit of a better idea. How is the quality of R&D to be judged? A good R&D manager or mentor will know but may find it hard to articulate. Proposal review panel members will find it more difficult because of their distance from the process.

A good operational definition of an exceptional R&D person is that we don't know what she or he will produce, but, whatever it is, it will be useful and more than worth the small investment. In other words, R&D is judged by results, not predetermined deliverables. These results must be relevant to the source of support, but they are not necessarily aimed at a specific scientific requirement. A good technology researcher knows the long-term goals of the science well enough to develop devices and techniques before most scientists know they need them.

## We need time to learn from new instruments and new technology

The full potential of new instruments takes considerable time to develop. Wider bandwidths present big challenges to data processing algorithms for accommodating frequency-dependent primary beam size, shape, and rotation, for example. The process of learning how to take advantage of new instrumentation invariably suggests improvements that were not foreseen in the original design. Larger, more complex instruments are more difficult to modify and take a longer time to respond to the learning process. New instruments built before their predecessors have run their course will make some of the same mistakes. All new technology and improvements in data analysis and observational strategies results from a clear understanding of past experience. This takes time.

Bigger does not necessarily mean more complex, and the reverse is also true. Some forms of complexity can be encapsulated and confidently retired. Very large scale integrated (VLSI) circuits are a good examples. Systems, such as large antenna structures, may be well understood from previous experience so their risk and cost uncertainty are relatively small. Software has proven much more problematic, and attempts to hide its complexity have met with very limited success. Even more unsettling is the fact that there is no common agreement on why this is the case. The jury is still out on firmware, the programming code that resides in FPGA's. The versatility of FPGAs for high speed signal processing is widely touted, but this could cause firmware to elude cost control, much like software. If firmware modules prove nearly as stable as VLSI chips, there's reason for optimism, but that risk has yet to be retired.

This white paper does not argue against large-scale radio astronomy instrumentation as such. We do offer a strong caution against complexity in the name of shared resources. New sophistication takes time to assimilate. Cost and schedule overruns and de-scoping experience with recent projects must be taken very seriously and the lessons thoroughly understood before attempting new large endeavors.

## Related reading

Three articles in the November 2008 issue of IEEE Spectrum Magazine on weapons procurement describe some uncomfortable parallels with the specification, design, development, and funding process being employed in the coming generation of large radio astronomy instruments. Some useful cautions can be found in these articles.
http://www.spectrum.ieee.org/nov08/6931 http://www.spectrum.ieee.org/nov08/6934
http://www.spectrum.ieee.org/nov08/6935